\documentclass[11pt]{article}
\usepackage{graphicx}
\usepackage{dcolumn}
\usepackage{bm}
\usepackage{float}
\usepackage{placeins}
\usepackage{hyperref}
\usepackage{natbib}

\usepackage{subfiles}

\usepackage[T1]{fontenc}
\usepackage[utf8]{inputenc}
\usepackage{authblk}

\title{Fermi-GBM Observation of GRB 090717034: $\chi^2$ Test Confirms Evidence of Gravitational Lensing by a Supermassive Black Hole with Million Solar Mass}
\author[1]{Zeinab Kalantari\thanks{zeinab.kalantari@physics.sharif.edu}}
\author[1]{Sohrab Rahvar}
\author[2]{Alaa Ibrahim}

\affil[1]{Department of Physics, Sharif University of Technology, Tehran 11365-9161, Iran}
\affil[2]{Department of Physics, College of Science, P.O.Box 36, P.C. 123, Muscat, Sultanate of Oman}

\begin{document}

\maketitle

\begin{abstract}

Gravitational lensing of gamma-ray bursts (GRBs) can provide an opportunity to probe the massive compact objects in the universe at different redshifts. We have discovered two consecutive pulses in the light curve of GRB 090717034, with the same temporal profile and different count rate, separated by a time interval, which is identified as gravitationally lensed candidate in Fermi/GBM GRB catalogue \citep{Kalantari}. Here, we use the $\chi^2$ method to investigate the similarity of the temporal profile variability of the two pulses as a gravitationally lensed candidate GRB. We find the magnification factor and the time delay between two pulses to correspond to minimising the $\chi^2$ function. Then, we perform a Monte Carlo simulation on a sample of mock lensed GRBs and  
compare the $\chi^2$ of the lensed GRB candidate with the simulation, which confirms this candidate with $1\sigma$ confidence level. Assuming that GRB 090717034 is lensed by a point-like object, the redshifted lens mass is about $M_L(1+z)=(4.220\pm 6.615) \times 10^6 M_{\odot}$. The lens of this GRB is a candidate for a super-massive black hole along the line of sight to the GRB.

\begin{description}
\item[Keywords]
Gamma Ray Bursts, Gravitational lensing, Black holes
\end{description}
\end{abstract}

\section*{}
The first GRB was detected in 1967 by the Vela satellite of the United States  \citep{Vela}. The Burst and Transient Source Experiment (BATSE) was then launched in 1991 \citep{BATSE} to study the origin of GRBs. This experiment has detected over 3000 GRBs and recognized an isotropic distribution of GRBs in sky \citep{BATSEIsotrop}. The spectroscopic measurements of the redshift of the host galaxies showed that GRBs are at cosmological distances from Earth \citep{GRBOrigin}. Our knowledge of GRBs has increased with the launch of the Swift and Fermi gamma-ray observatories in 2004 and 2008, respectively \citep{swift,MeeganGBM}. Their observations revealed new populations of GRBs, such as GRBs at high redshift \citep{highZgrb}, ultra-long GRBs, and ultra-high-energy (GeV) GRBs \citep{highEgrb}. The Gamma-ray Burst Monitor (GBM) on board of the Fermi Gamma-ray Space Telescope (Fermi-GBM) records GRBs in a broad energy range (8-40 MeV) and with a field of view of $\geq8$ sr. This instrument has observed more than 2000 long GRBs (with $T_{90} \geq 2$s) and is ideal for finding gravitationally lensed GRBs \citep{MeeganGBM, von_Kienlin_2020}.
\par
GRBs are ideal sources to probe the cosmos due to their high redshifts. Gravitational lensing of GRBs has been studied for many ars. \cite{PaczynskiLensedGRB} was the first to propose the idea that the GRBs could be gravitationally lensed. Null results have been obtained in searches for strong gravitationally lensed GRBs, where the time delay between the images is greater than the duration of the burst \citep{Marani1996,MaraniLensedGRB,DavidsonLensedGRB,LiLensedGRB,HurleyLensedGRB,AhlgrenLensedGRB}. On the other hand, some gravitationally microlensed GRB candidates have been introduced by investigating their light curves \citep{OugolnikovMicroGRB,HiroseMicroGRB,PaynterMicroGRB,Kalantari,Yang_MicroGRB,VeresMicroGRB}.
\par
In our recent work, we applied the autocorrelation technique to the long GRB light curves within the ars of 2008-2020 in the Fermi-GBM catalog and found one candidate (GRB 090717034) as a lensed GRB with a redshifted lens mass of $M_L(1+z)\simeq 10^6 M_\odot$ for a point-mass model of the lens \citep{Kalantari}. This lens induced a time delay of $\approx 41.332 $s and a flux ratio of about $1.782$ between the two GRB images. This candidate passed two criteria which verify that the two pulses of the GRB candidates are real and not artifacts. As a first step, we select GRB candidates that show two distinct pulses in their light curves, with a faint pulse following a bright pulse in the energy range beyond which the GRB spectrum does not exhibit overflow or cutoff effects, which usually is between 8 and 150 keV. Secondly, we choose the NaI detector that shows the brightest signal, and to avoid any instrumental error, we also checked that at least three other NaI detectors recorded the two pulses for the selected GRB in different energy bands, therefore, GRB 090717034 has been regarded as a candidate for gravitational micolensing \citep{Kalantari}. \cite{Nemiroff1,Nemiroff2} challenged this candidate, arguing that the pulse shapes are different at about the $3.1 \sigma$ confidence level. But \cite{Lin_responses} by repeating the same $\chi^2$ test mentioned by \cite{Nemiroff2} on the light curve of GRB 090717034 in a different time bin concluded the result of \cite{Nemiroff2} $\chi^2$ test is very sensitive to the time bin sizes of the light curve. \cite{Lin_responses} confirmed GRB 090717034 as the first-class candidate of gravitational lensing based on its excellent results in the auto-correlation test, hardness test, and time-integrated/resolved spectrum test  
In this paper, we revisit the light curve of GRB 090717034 and use 
the $\chi^2$ test to examine the credibility of this event as a lensed GRB candidate.

\begin{figure}[!htbp]
\centering
\includegraphics[width=.7\linewidth]{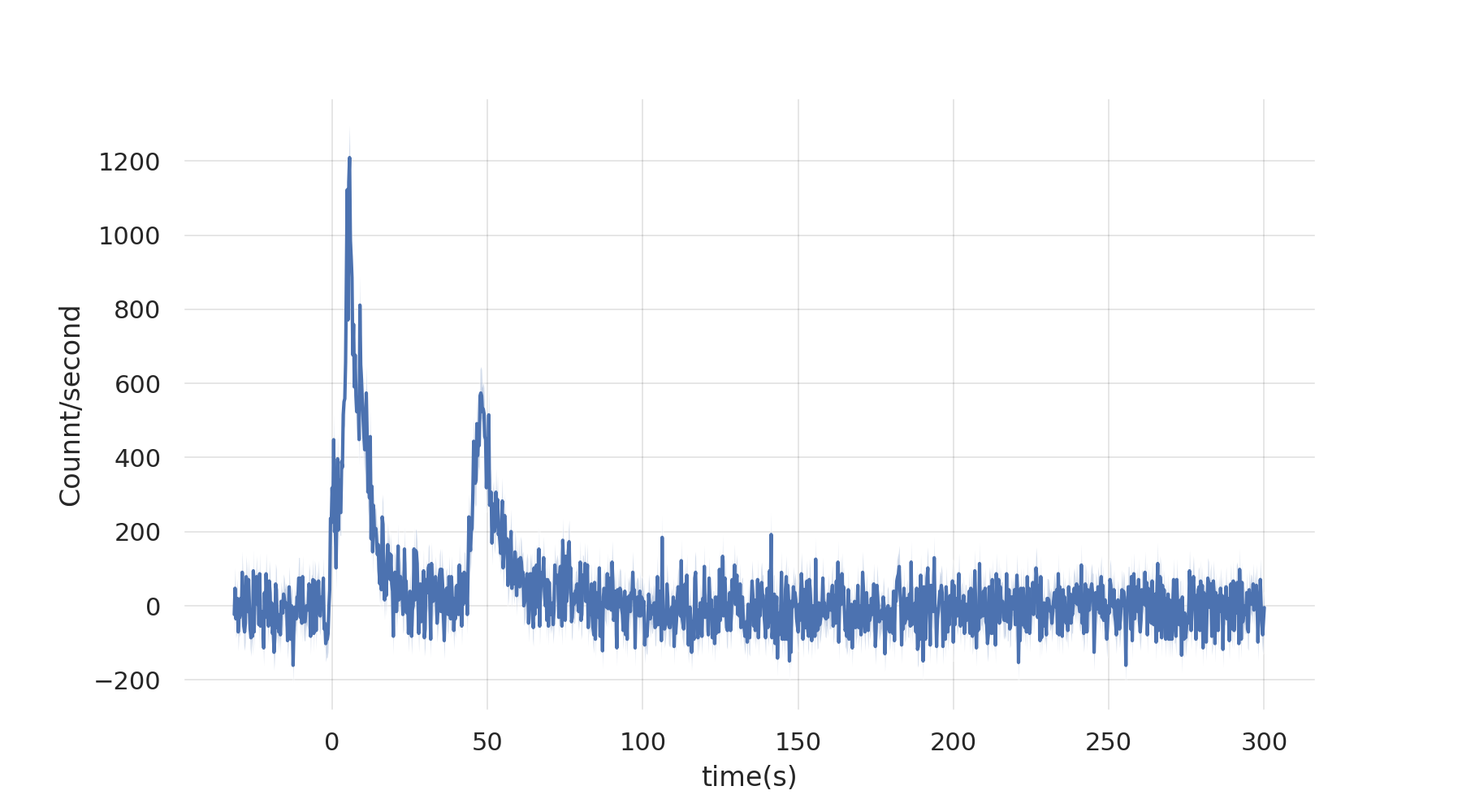}

\caption{ The light curve of GRB 090717034 (with Fermi-GBM) from NaI 1 detector with time bin of 256 ms. This event fulfills the two criteria we considered to be a gravitationally lensed candidate (See text above and \cite{Kalantari}). 
}
\label{fig:LCGRB}
\end{figure}

In \cite{Kalantari}, we investigated the possiblity of  gravitational lensing of GRB in the Fermi/GBM data and found one candidate among 2137 long GRBs searched. We extracted light curves from the NaI detectors that record the highest signal and used the time-tagged event (TTE) file for GRB 090717034 in the energy range of $8-150$ keV (where the GRB spectrum did not show overflow or cutoff effect). The light curve of GRB 090717034 is depicted in Fig.\ref{fig:LCGRB}. In our previous data analysis, the autocorrelation method gave the time delay between the two pulses and the magnification ratio after the subtraction of the background. According to the light curve of this GRB, a zeroth polynomial model (a constant) is chosen to fit the background. The two physical parameters from the autocorrelation method were reported as $\tau = 41.332\pm 0.283$s (time delay between the pulses) and $R= 1.782\pm 0.239$ (as the magnification ratio).

\par
Here we repeat our analysis using the $\chi^2$ method. After subtracting the background, we shift back the second pulse by the amount $\tau$ and magnify it with the factor $R$ to superimpose the two pulses. We take into account the error in count rate as a Poisson noise which is proportional to the square root of the photons count in each time bin ($\sqrt N$). Then we compare the two light curves using the $\chi^2 $ function \citep{Chi2}
 
\begin{equation}
\chi^2(\tau,R) = \frac{1}{N_I-2}\sum_{t=t_{start}}^{t_{start}+T_{90}/2} \frac{(I_1 (t)-R \times I_2(t-\tau))^2}{\sigma_1(t)^2+R\times \sigma_2(t-\tau)^2} 
    \label{Eq:chi}
\end{equation}

Here $I_1(t)$, $\sigma_1(t)$ and $I_2(t)$, $\sigma_2(t)$ respectively are the count rate and the error in the count rate of the first and second pulses in the GRB light curve and $N_I$ is number of data bin in the range of summation and $\chi^2$ is normalized to the number of degrees of freedom. $T_{90}$ is a measurement of the time interval in which 90\% of the total observed counts have been detected and $t_{start}$ is defined by the time at which 5\% of the total counts have been detected. Since gravitational lensing does not change the time duration of two images, we assume that the duration of each pulse of the light curve (images) is about half of $T_{90}$. Therefore we calculate the $\chi^2$ of the gravitationally lensed candidate in the time interval of $t_{start}$ to $t_{start}+T_{90}/2$ . We used data from the Fourth Fermi/GBM Gamma-Ray Burst Catalog \citep{von_Kienlin_2020} of GRB 090717034 for the $T_{90}$ duration and the start time of this burst. 


\par
\begin{figure}[!htbp]
\centering
\includegraphics[width=1\linewidth]{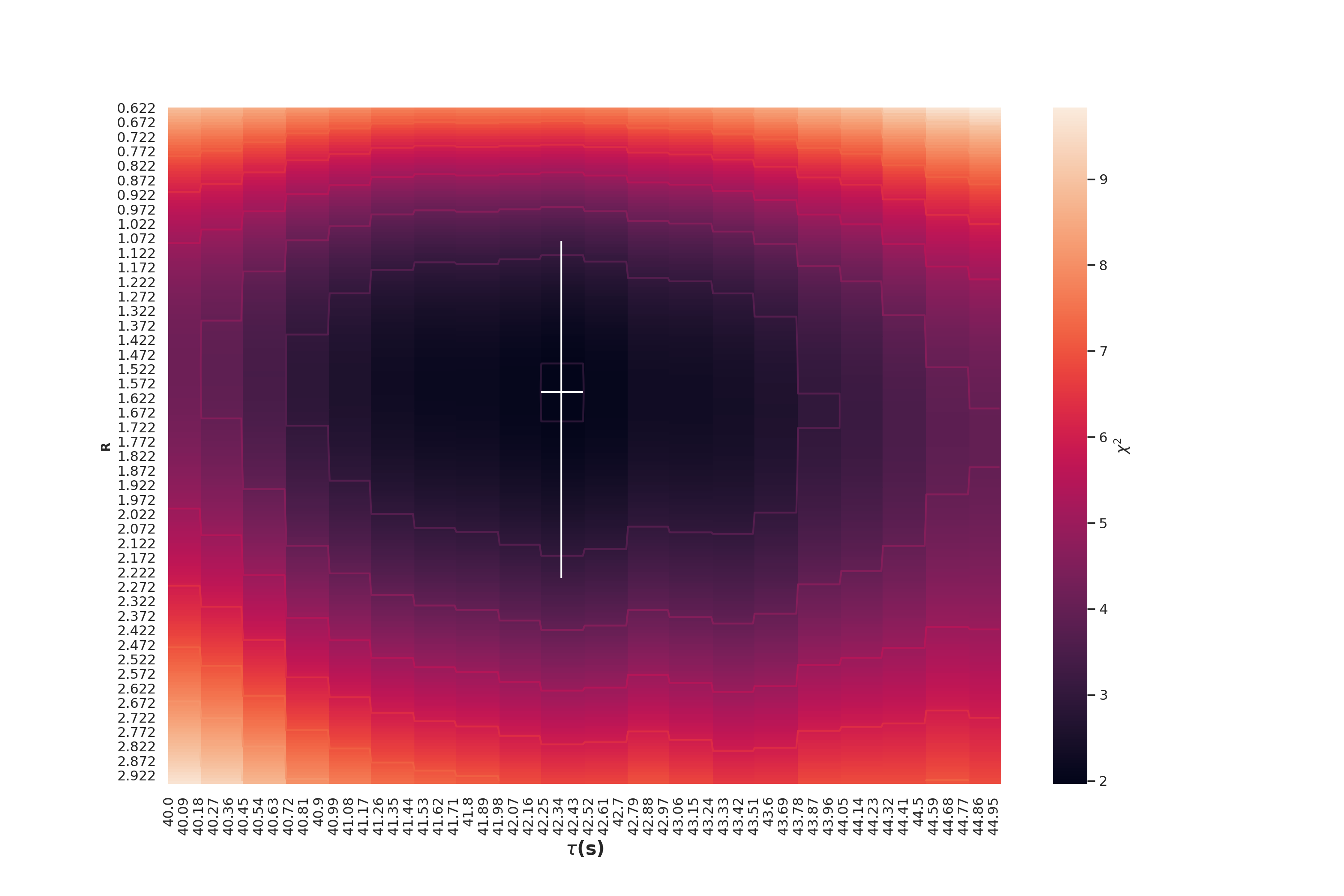}
\caption{\label{fig:chicount} Normalized $\chi^2$ of GRB 090717034 obteined from comparing the light curves of the two pulses of the GRB by uising a range value for the magnification factor $R$ and time delay $\tau$ (Eq.(\ref{Eq:chi})). The minimum of $\chi^2$ is $1.961 \pm 1.337$ occurs at $R=1.662 \pm 0.580$ and $\tau= 42.365 \pm 0.125 $s considering $1\sigma$ confidence level in $\chi^2$ depicted by the white cross sign.  }
\end{figure}

We calculate the $\chi^2$ of the lensed candidate for a range of parameters of $R$ and $\tau$ around the parameters value that we found from autocorrlation method ($0.2 R_{ACF}\le R \le 2R_{ACF}$) and $0.9 \tau_{ACF}\le \tau \le 1.1\tau_{ACF}$). 
The minimum value of $\chi^2$ is found $1.961\pm 1.337$that is occurring at $R=1.662 \pm 0.580$ and $\tau= 42.365 \pm 0.125 s$ with $1\sigma$ error in $\chi^2$. Fig.\ref{fig:chicount} depicts the value of $\chi^2$ in terms of two parameters $R$ and $\tau$. 

\begin{figure}[!htbp]
\centering
\includegraphics[width=0.7\linewidth]{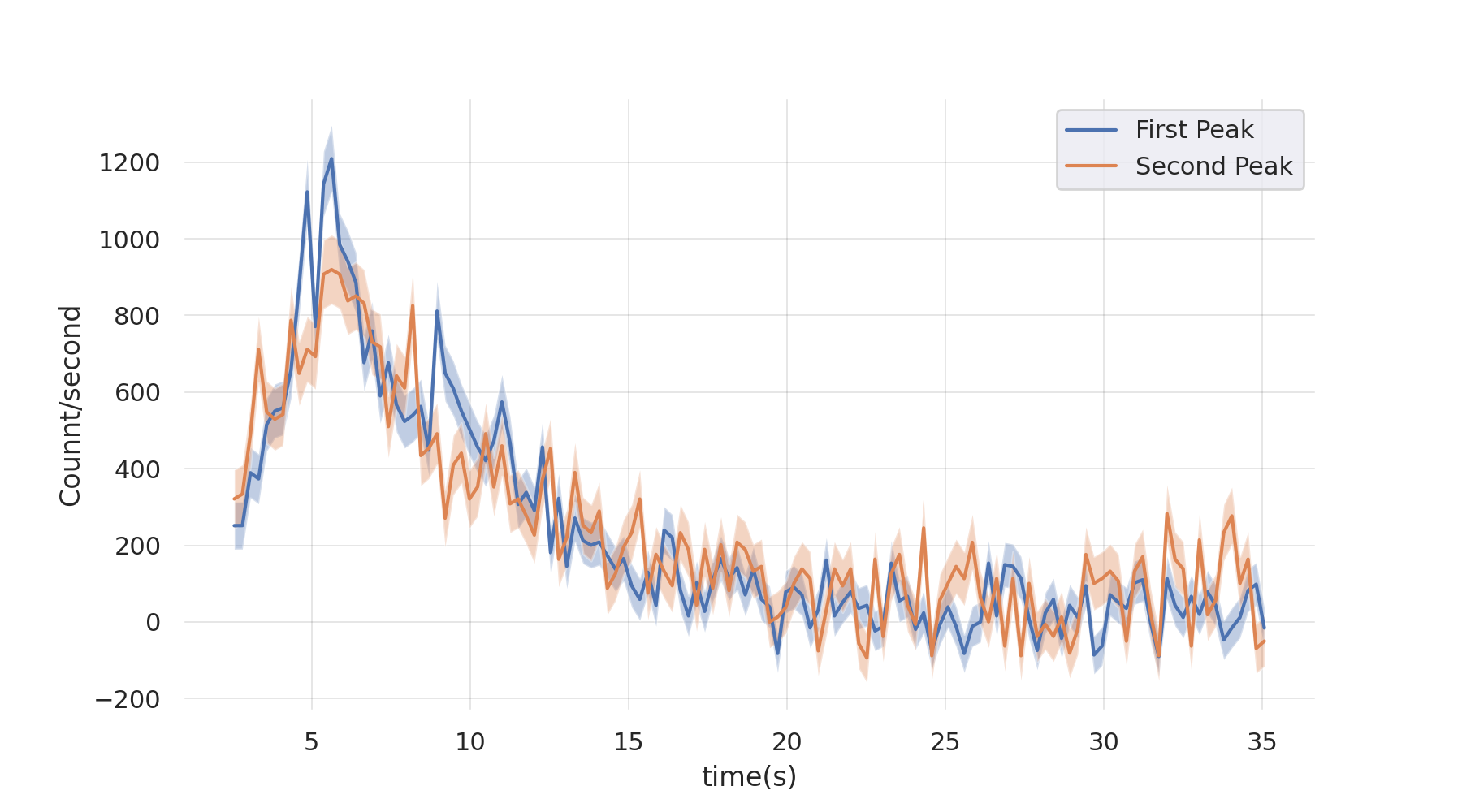}
\caption{\label{fig:LC_2pulses} Comparing the two pulses of GRB 090717034 after the fainter pulse is magnified with the amount $R=1.662$ and shifted back by $\tau = 42.365 s$. }
\end{figure}

 As a visual check of the validity of the best values the parameter of $R$ and $\tau$, after subtracting the background from the light curve of the lensed candidate, we magnify the second pulse and shift it with the amount of the time delay. Fig. (\ref{fig:LC_2pulses}) shows the overlap of the two pulses in the burst duration ($T_{90}/2$) with the time delay and the magnification factor that were found from the $\chi^2$ minimization method. 

We also checked the reliability of $\chi^2$ fitting by performing a Monte Carlo simulation of mock lensed GRBs. For this simulation, we select 81 long GRBs from the Fermi/GBM catalog \citep{von_Kienlin_2020} that have one pulse in their light curves with time bin of 64 ms. We produced mock lensed light curves of these GRBs by imposing a time delay and a magnification factor that we found from the $\chi^2$ method ($\tau = 42.365 $s and $R= 1.602$). Then we find the $\chi^2$ value for each of these mock lensed data in the time range of their $t_{start}$ to $t_{start}+T_{90}$. The mean value of $\chi^2 $ in the simulation was $3.309 \pm 5.436$. Therefore, the $\chi^2$ test of our lensed candidate verified that two pulses of this GRB light curve are intrinsic and with high confidence level as due to gravitational lensing. The lens mass for a point-mass model is derived to be $M_L(1+z)=(4.220\pm 6.615) \times 10^6 M_{\odot}$ for the time delay and magnification factor based on the result of the $\chi^2$ method. Table 1 compares the best values for the parameters of the lens from the $\chi^2$ and autocorrelation function methods. We note that gravitational lensing of GRBs can identify the compact and non-compact structure in the universe.
The time delay between the pulses reaching the observer is an indication of the compactness of the lens. The diffused structures as galaxies and globular cluster may cause time delay, in the images in the order of months and days, however, for compact lenses as black holes, the time delay is in the order of the time needed to cross the horizon of the black hole (i.e. $\tau = R_{s}/c$). 


\begin{table*}[!htbp]

\caption{\bf Compering the magnification and the time delay of GRB 090717034  with two techniques}
\begin{tabular}{ccccc}
Method &$\tau$ &  $R$ & $M_L(1+z_L)$ \\
\hline
Autocorrelation & $41.332\pm 0.283$ s & $1.782\pm 0.239$ & $(3.615\pm
1.763)\times10^{6}M_\odot$\\
\hline
Minimising $\chi^2$ & $42.365 \pm 0.125$ s & $1.662 \pm 0.580$ & $(4.220\pm 6.615)\times10^{6}M_\odot$\\
\end{tabular}
\label{tab:Compering}
\end{table*}

Concluding this work, we used the $\chi^2$ test on the light curve of GRB 090717034 from Fermi-GBM observations \citep{Kalantari} to examine this event as a gravitationally lensed candidate. We found the best values of the  magnification ratio and time separation of two local pulses in the GRB light curve to be consistent with those obtained from the autocorrelation method. Our result for magnification factor and time delay are updated to $R=1.662 \pm 0.580$ and $\tau= 42.365 \pm 0.125 $s. They are within  7\% and 3\%, respectively from the values obtained from the autocorrelation technique (table \ref{tab:Compering}). We also found the acceptable value for best $\chi^2$ by performing a Monte Carlo simulation on a sample of 81 mock gravitationally lensed data with the same magnification and time delay of the  gravitationally lensed candidate.  The best value for $\chi^2$ of GRB 090717034 light curve is $1.961 \pm 1.337$  which is well within the acceptable range of Monte Carlo simulation with the mean value of $3.309 \pm 5.436$. This comparison confirms that this event has a true lensing signature and is not a result of stochastic random fluctuations in the GRB light curve.

Under assumption of point mass-model for the gravitational lens, the lens mass was derived as $M_L(1+z)=(4.220\pm 6.615) \times 10^6 M_{\odot}$, which confirmed a signature of a supermassive black hole. One of the scenarios for the formation of supermassive black holes (SMBHs) are thought to form from primordial black holes (PBHs) with masses of $10^6 M_{\odot}$. The PBH concept was first developed in 1974 by \cite{Hawking_PBH} and they suggest that over-dense regions of the early universe could collapse into black holes. Different authors have proposed various mechanisms to explain this phenomenon\citep{Suyama_PBH,Jedamzik_PBH,Bullock_PBH}. One plausible hypothesis is that PBHs are formed as a result of large-scale density fluctuations in the early universe \citep{Garc_PBH}.A variety of observations constrained the proportion of dark matter in the form of PBHs in different mass ranges \citep{Ha_DM,Carr_DM,Chen_DM,Wang_DM,Wang_DM2,Aloni_DM,Magee_DM,Abbott_DM,Graham_DM,Griest_DM,Nakama_DM,Gaggero_DM,Niikura_DM,Zumalac_DM,Barnacka_DM,Zu-Cheng_DM,Tisserand_DM,Abbott_Abbott_DM,Cappelluti_CITE,2021MNRAS.507..914R,2021PhRvD.103h4001K}. 

The study of gravitational lensing of GRBs is an important tool to constrain the mass fraction of PBH to the total mass of dark matter in the windows where PBHs is not t constrained. In order to achieve this goal, we need GRB data with higher accuracy in photon counting as well as better sensitivity in identifying the poisons of GRBs. Rich statistics from the lensed GRBs with control on the bias of the detectors may answer the challenging question on dark matter: whether dark matter is formed from the large mass primordial black holes? 



\bibliographystyle{unsrtnat}
\bibliography{References}



\end{document}